\documentclass[usenatbib]{mn2e}
\usepackage{multirow}
\usepackage{epsfig,amsmath,amssymb,aas_macros,bm}

\newcommand{\solarmass}{$h^{-1} \rm M_\odot$}

\newcommand{\mpch}{$h^{-1} \rm Mpc$~}
\newcommand{\msub}{$m_{\rm sub}/M_{200}$~} 

\newcommand{\mmsub}{m_{\rm sub}/M_{200}~} 
\newcommand{\muone}{\widetilde{\mu_1}}
\def\gsim { \lower .75ex \hbox{$\sim$} \llap{\raise .27ex \hbox{$>$}}}
\def\lsim { \lower .75ex \hbox{$\sim$} \llap{\raise .27ex \hbox{$<$}}}
\title[LCDM subhaloes]
{The statistics of the subhalo abundance of dark matter haloes}
\author[Gao et al.]
       {L. Gao$^{1,2,}$\thanks{Email:lgao@bao.ac.cn},
        C. S. Frenk$^2$, 
        M. Boylan-Kolchin$^3$, 
        A. Jenkins$^2$, 
        V. Springel$^{3,4}$, 
        \newauthor
        S. D. M. White$^3$ \\        
$^1$National Astronomical Observatories, Chinese Academy of Science,
Beijing, 100012, China \\
$^2$Institute of Computational Cosmology, Department of Physics,
University of Durham,\\ Science Laboratories, South Road, Durham DH1
3LE \\
$^3$Max-Planck Institute for Astrophysics, Karl-Schwarzschild Str. 1,
D-85748, Garching, Germany \\
$^4$Heidelberg Institute for Theoretical Studies, Schloss-Wolfsbrunnenweg 35,
69118 Heidelberg, Germany \\
}

\begin{document}
\label{firstpage} \maketitle
\title{The statistics of LCDM subhaloes}

\begin{abstract}
  We study the population statistics of the surviving subhaloes of $\Lambda$CDM
  dark matter haloes using a set of very high resolution $N$-body
  simulations. These include both simulations of representative regions of the
  Universe and ultra-high resolution resimulations of individual dark matter
  haloes. We find that more massive haloes tend to have a larger mass fraction
  in subhaloes. For example, cluster size haloes typically have 7.5 percent of
  the mass within $R_{200}$ in substructures of fractional mass larger
  than $10^{-5}$, which is $25$ percent higher than galactic
  haloes. There is, however, a large variance in the subhalo mass
  fraction from halo to halo, whereas the subhalo abundance shows much
  higher regularity. For dark matter haloes of fixed mass, the subhalo
  abundance decreases by $30$ percent between redshift $2$ and $0$.
  The subhalo abundance function correlates with the host halo
  concentration parameter and formation redshift.  However, the
  intrinsic scatter is not  significantly reduced for narrow ranges of
  concentration parameter or  formation redshift,  showing that they
  are not the dominant parameters that determine the subhalo abundance
  in a halo.
\end{abstract}


\section{Introduction}

In the standard $\Lambda$CDM model of cosmogony, cold dark matter
haloes form from primordial Gaussian fluctuations by accretion of
diffuse dark matter and mergers with other haloes. During this
hierarchical process, accreted haloes often survive as self-bound
substructures or subhaloes orbiting around the main
system~\citep[e.g.,][]{T98, ghigna00, springel01, delucia04, g04a,
diemand04, reed05, diemand07, springel08a}. The most massive subhaloes are
likely hosts of luminous satellite galaxies.

The evolution of subhaloes is determined by the interplay of phenomena
such as accretion, mergers, dynamical friction and tidal
stripping. Thus, subhalo properties are most directly and accurately
investigated using simulations. So long as one is exclusively
interested in evolution of the dark matter, this is a purely
gravitational problem that is ideally suited to N-body
techniques. Indeed, most work on subhaloes has exploited this
approach. However, since subhaloes represent a small fraction of the
total mass of the halo, most studies have been based, for
computational resource reasons, on high resolution simulations of a
handful of individual haloes~\citep[e.g.,][]{ghigna00, springel01,
stoehr03, g04a, diemand04, diemand07, springel08a}.

Statistical studies are much more demanding because of the need to simulate
large volumes with a large dynamic range and, to date, have focused on
subhaloes in relative massive haloes~\citep[e.g.,][]{delucia04,g04a,k04,shaw06,
  elahi09, angulo09}. Inspired by, and calibrated from, the results of
simulations, semi-analytic models have been developed to calculate certain
properties of subhalo populations~\citep[e.g.,][]{benson05, tb05, van05,
  zentner05,giocoli08a, giocoli08b, lm09}.

In this short paper, we exploit the unprecedented statistical power
and high resolution of the {\it Millennium-II Simulation} \citep{bk09},
together with other recent cosmological simulations and a suite of
very high resolution resimulations of individual galaxy haloes
\citep{springel08a} to study the abundance of subhaloes with much
better statistics and over a significantly larger dynamic range than
has been possible in the past. We address three specific questions
concerning the subhalo mass function: (i) how does it depend on the
mass of the parent halo? (ii) how does it depend on redshift, for a
fixed halo mass? (iii) how does it depend, at fixed mass and redshift,
on the properties of the halo, specifically the concentration and
formation redshift? Some of these questions have been addressed
in some of the earlier studies mentioned above. Our results extend
these studies into a regime of subhalo mass not previously probed and
with much greater statistical power.

\section{Numerical simulations}

The main advance in this paper comes from analysing two sets of very
recent, high resolution simulations. One is the 10-billion particle
Millennium-II Simulation~({\small MS-II}, \citealt{bk09}) of a
100\mpch cubic volume. The other one is the set of 6 resimulations in
the Aquarius Project~\citep{springel08a,springel08b}, each of
which follow the evolution of a galaxy size halo with more than
$10^{8}$ particles inside the virial radius. All these simulations
represent a significant improvement over the previous generation of
N-body simulations. We supplement these data with older simulations,
the original 10-billion particle Millennium Simulation~({\small MS}),
\citealt{springel05b}) of a 500\mpch cubic volume and the h{\small MS}
simulation~\citep{g08},which has matching phases in the initial
conditions to those of the {\small MS-II}, but with about $14$ times
lower mass resolution and $2.4$ times lower spatial resolution.

All the simulations assume the same values of the cosmological
parameters: mean matter density, $\Omega_m=0.25$; cosmological
constant term, $\Omega_{\Lambda}=0.75$; Hubble parameter (in units of 100~${\rm
  km}\,{\rm s}^{-1}\,{\rm  Mpc}^{-1}$), $h=0.73$;
fluctuation amplitude, $\sigma_8=0.9$; and primordial spectrum
power-law index, $n=1$; These values are consistent with the first
year {\small WMAP} data~\citep{spergel03} but differ at about the
$2-\sigma$ level from more recent determinations. This small off-set
is of no consequence for the topics addressed in this paper. All the
simulations were run with either the Gadget-2~\citep{springel05a} or
the more recent, Gadget-3, code used for the Aquarius Project.

The parameters of the simulations are summarized in
Table~\ref{tab:sims}. Note that the {\small MS} and the {\small MS-II}
differ in mass resolution by a factor of 125, while the {\small MS}
and the h{\small MS} differ by a factor of 9, allowing numerical
convergence tests to be carried out.

\begin{table}
\caption{Simulation parameters: (1) name of the simulation; (2) the
  side of the simulated cubic volume; (3) the number of particles in
  the simulation; (4) the mass per particle; (5) the gravitational
  softening length. The initial conditions for h{\small MS} and
  {\small MS-II} simulations have matching phases. 
}
\begin{tabular}{lccccc}
\hline
Name & Box size & $N_{\rm p}$ &${m_{\rm p}}$ & $\epsilon$\\
&  [$h^{-1}\,$Mpc] &  & [$h^{-1}\,{\rm M_{\odot}}$] &
[$h^{-1}\,$kpc]\\
\hline
\hline
{\small MS} &$500$ &$2160^3$ &$8.6 \times 10^{8}$ &$5.0$ \\
\hline
h{\small MS} &$100$ &$900^3$ &$9.5 \times 10^{7}$ &$2.4$ \\
\hline
{\small MS-II} &$100$ &$2160^3$ &$6.9 \times 10^{6}$ &$1.0$ \\
\hline
{Aquarius} & \multirow{2}{*}{--}& \multirow{2}{*}{--} &\multirow{2}{*}{$\sim 1 \times 10^{4}$} &\multirow{2}{*}{0.048}\\
(level 2) & & & &\\
\hline
\end{tabular}

\label{tab:sims}
\end{table}

Dark matter haloes were found using the friends-of-friends
algorithm~\citep{davis85}, and their self-bound subhaloes were identified
using the {\small SUBFIND} algorithm~\citep{springel01}. Haloes and
their subhaloes were tracked across the simulation outputs and linked
together in merger trees. In what follows, we restrict our attention to
subhaloes contained within the virial radius, ${R_{\rm 200}}$, of
their parent host halo, defined as the radius within which the
enclosed mass density is $200$ times the critical value. We note that
\cite{mike} used a different definition of the virial radius, based
on the spherical collapse model~\citep{Eke96}.

\section{Results}
\begin{figure} 
\resizebox{9cm}{!}{\includegraphics{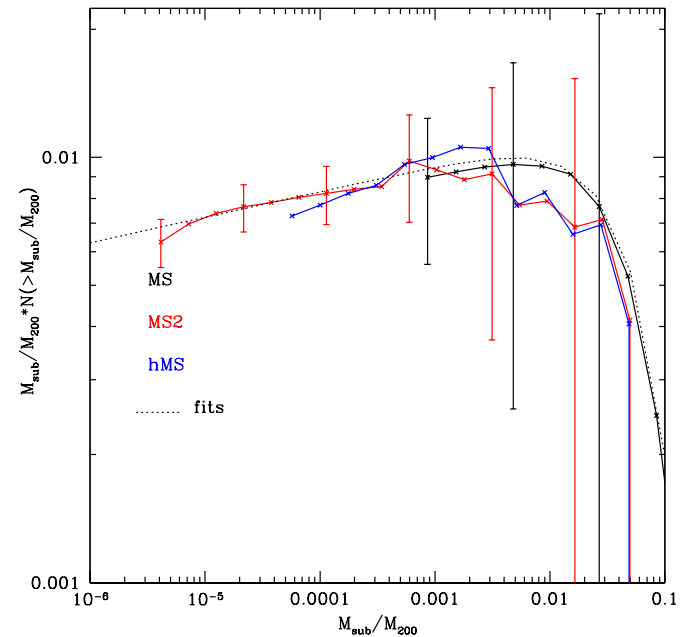}}
\caption{The mean cumulative mass function of subhaloes in hosts with masses in
  the range $[1,3] \times 10^{14}$\solarmass, as a function of subhalo mass
  fraction, $M_{\rm sub}/M_{200}$. The $y$-axis has been multiplied by the mass
  fraction in order to expand the dynamic range. The 3 solid lines show the
  averaged subhalo mass function for samples of 12 haloes in the {\small MS-II}
  (red) and the h{\small MS} (blue) and of 1676 haloes in the {\small MS}
  (black), as indicated in the legend. An analytic fit to the {\small MS} data
  using eqn.~(\ref{eq:fits}) is shown as a dotted line.  The error bars on
  selected points show the {\it rms} scatter about the mean for the {\small MS}
  and {\small MS-II} haloes.}
\label{fig:conv}
\end{figure}

We start by considering the cumulative subhalo mass functions for
samples of haloes with mass in the range $[1, 3] \times
10^{14}$\solarmass\ in the {\small MS-II}, h{\small MS} and {\small
MS} simulations at $z=0$. The first two samples comprise the same 12
haloes at different resolutions, while the third contains 1676 haloes.

The mean cumulative mass functions of subhaloes for all three samples are
plotted in Fig.~\ref{fig:conv}. Here, and below, we show cumulative mass
functions multiplied by $M_{\rm sub}/M_{200}$ in order to take out the dominant
mass dependence and make the differences between curves more
apparent. Following previous work ~\citep[e.g.,][]{stoehr03, g04a,
  diemand07, springel08a}, we restrict the samples to include only
subhaloes with 100 particles or more in order to avoid numerical
effects. There is good convergence between the results from the
{\small MS-II} and the h{\small MS} and, within the errors, there is
good agreement with the cumulative mass function of the {\small
  MS}. The error bars show the rms scatter about the mean for the MS
and MS-II haloes. (They are plotted at a selection of well-separated
points to reduce the effect of correlations between adjacent bins.)
This scatter, which reflects the varied formation histories of haloes,
is substantial, particularly for large values of $M_{\rm sub}/M_{200}$.

In a recent study, \cite{mike} demonstrated that the subhalo
occupation distribution is well described by a negative binomial,
independently of host halo mass. In this case, the scatter at the high
mass end of the subhalo mass function is well described by a Poisson
distribution, but at the low mass end, \msub$<10^{-4}$, 
intrinsic scatter of fractional amplitude $\sigma \sim 0.15$ dominates.
The authors further found that the cumulative subhalo mass function of galactic
size haloes can be well fitted with the following form:

\begin{equation}\label{eq:fits}
  N(>\mu \equiv \mmsub) = \left(\frac{\mu}{\muone} \right)^a \, \exp \left[-\left(
      \frac{\mu}{\mu_{\rm cut}} \right)^b    \right] \,.
\end{equation}

The dashed line in Fig.~\ref{fig:conv} is a fit to the mean subhalo cumulative
mass function for the cluster size haloes from the {\small MS} using
eqn.~(\ref{eq:fits}) with parameters $\muone=0.01$, ${\mu_{\rm cut} =0.1}$,
$b=1.2$ and power-law index $a=-0.94$.  Clearly, the fit is very good.  The
best-fit parameters differ slightly from the values given by \cite{mike}
because a different definition of virial radius has been used here and the
range of halo masses is different.

\subsection{Dependence on host halo mass}
\begin{figure} 
\resizebox{9cm}{!}{\includegraphics{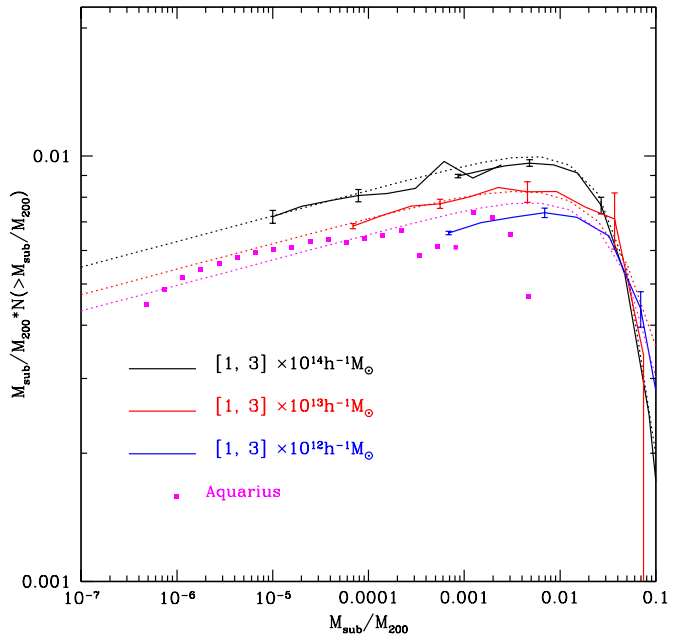}}
\caption{The dependence of the subhalo mass function on the host halo mass. The
  solid lines show the averaged cumulative subhalo mass functions for three
  intervals of host halo mass: $[1,3] \times 10^{14}$ \solarmass\ (black),
  $[1,3] \times 10^{13}$\solarmass\ (red) and $[1,3] \times
  10^{12}$\solarmass\ (blue). Haloes in the first mass range come from the
  {\small MS} and {\small MS-II}, while those in the latter two ranges come
  from the {\small MS-II}. The error bars on selected points show the error on
  the mean for the three mass ranges indicated. The filled squares show the
  mean of the cumulative subhalo mass functions of the 6 Aquarius
  haloes. The dotted lines show fits to eqn.~(\ref{eq:fits}); the fit
  parameters are listed in eqn.~(\ref{eq:fitmass}). }
\label{fig:massdep}
\end{figure}

\cite{g04a} noted that the abundance of relatively
massive subhaloes (${\rm M_{sub}/M_{200}} > 0.001$) increases
systematically with host halo mass. This trend reflects the fact that
in the CDM cosmology more massive haloes are, in the mean, both less
centrally concentrated and younger than less massive haloes. Thus,
they exert weaker tidal forces and have had less time to disrupt
their substructure. This result has been confirmed by subsequent
numerical and semi-analytical
studies~\citep{zentner05,van05,shaw06,giocoli08b}.

The much larger samples of well resolved haloes in our set of
simulations allows us to explore the trend with host halo mass to much
smaller mass ratios than have been resolved in earlier simulations
which, for the most part, have focused on cluster size haloes
\citep[e.g.][]{g04a,zentner05}. Combining data from the Aquarius and
{\small MS-II} simulations, we have a sample of haloes spanning the
range $\sim 10^{12}- 3 \times 10^{14}$\solarmass, each resolved with
at least $10^5$ particles within the virial radius, and often with 10
to  1000 times more. This represents an improvement by a factor of
100-1000 over the simulations analyzed by~\cite{g04a}.

Fig.~\ref{fig:massdep} shows cumulative subhalo mass functions for
host haloes of different mass. The magenta squares and the solid lines
respectively represent galactic size haloes ($[1, 3]
\times 10^{12}$\solarmass) from Aquarius and
{\small MS-II}, the blue lines represent group size haloes ($[1, 3]
\times 10^{13}$\solarmass) from the {\small MS-II} and the black lines
represent cluster size haloes ($[1, 3] \times 10^{14}$\solarmass) from
the {\small MS-II} and the {\small MS}. The corresponding best fits of
the form of eqn.~(\ref{eq:fits}) are overplotted as dotted
lines. Clearly, eqn.~(\ref{eq:fits}) provides an excellent fit to the
cumulative subhalo mass function for all three ranges of host halo
mass, albeit with varying best-fit parameters.  We fixed the subhalo
function slope index to $a=-0.94$ when fitting all three curves.  Of
the remaining three free parameters in eqn.~(\ref{eq:fits}), $\muone$
determines the overall amplitude of the subhalo abundance function and
is largely independent of the parameters $\mu_{\rm cut}$ and $b$ which
combine to determine the shape of the cut-off at the high mass end. We
find that neither $\mu_{\rm cut}$ nor $b$ can be well constrained
on their own, but are largely degenerate.  Choosing $\mu_{\rm cut}$ in
the range $[0.04, 0.1]$ or $b$ in the range $[0.8,1.5]$ gives fits
which match the overall subhalo halo mass function reasonably well. To
break the degeneracy between $\mu_{\rm cut}$ and $b$ in our fits we
set $b = 1.2$ and fitted eqn.~(\ref{eq:fits}) by varying just $\muone$
and $\mu_{\rm cut}$. 
The best-fit values for $\muone$ and $\mu_{\rm
cut}$ are:
\begin{eqnarray}
M_{200} \in [1,3] \! \times \! 10^{14} \, h^{-1} \, {\rm M_{\odot}}\! : \;
\muone=0.0110,\; \mu_{\rm cut}=0.10. \nonumber \\
M_{200} \in [1,3] \! \times \! 10^{13} \, h^{-1} \, {\rm M_{\odot}}\! : \;
\muone=0.0092,\; \mu_{\rm cut}=0.07. \nonumber\\
M_{200} \in [1,3] \! \times \! 10^{12} \, h^{-1} \, {\rm M_{\odot}}\! : \;
\muone=0.0085,\; \mu_{\rm cut}=0.08. 
\label{eq:fitmass}
\end{eqnarray}

\begin{figure} 
\resizebox{9cm}{!}{\includegraphics{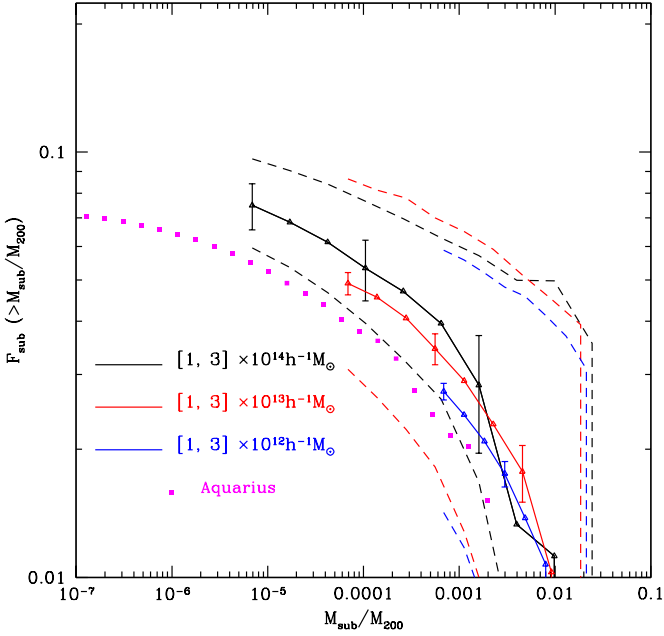}}
\caption{The mass fraction in substructures in haloes of different
mass as a function of the normalised subhalo mass. 
The lines show the medians of the cumulative subhalo mass
fractions for three ranges of host halo mass in the {\small MS-II}
simulation: $[1,3] \times 10^{14}$
\solarmass\ (black solid lines), $[1,3] \times 10^{13}$\solarmass\ 
(red solid lines) and $[1,3] \times 10^{12}$
\solarmass\ (blue solid lines). The magenta squares show 
the averaged subhalo mass fraction in the 6 Aquarius haloes. The
dashed lines show the 20 and 80\% of the corresponding
distribution. The error bars on selected points show the
error on the median.}
\label{fig:massfrac}
\end{figure}

By examining the subhalo mass functions of individual halos, we
have found that the main scatter is in the normalisation rather
than the shape; halos with a higher than average abundance of
low-mass subhalos also tend to have a higher than average abundance
of high-mass halos and vice versa.

At a given ${\rm M_{sub}/M_{200}}$, there is a weak trend in the
abundance of subhaloes with host halo mass. In the region of overlap in
${\rm M_{sub}/M_{200}}$, the cluster haloes have a mass fraction of
substructures that is about $25\%$ higher than in the galaxy haloes. The
mean of the group sample is intermediate between these two. The $15\%$
difference between group and cluster halo identified by \cite{g04a} is
visible at the smaller mass ratios plotted, ${\rm M_{sub}/M_{200}}
\sim 10^{-4}$.  The mass dependence of the subhalo abundance with host
halo mass in our simulations is much weaker than a theoretical
expectation based on semianalytical modelling of the subhalo
population~\citep[e.g.,][]{van05}.

The mass fractions in subhaloes as a function of relative mass for host haloes
of different size are shown in Fig.~\ref{fig:massfrac}. The curves are very
steep at high values reflecting the fact that most of a typical halo's mass in
subhaloes is contributed by a relatively small number of very massive
subhaloes.  As shown by \citet{springel08a} for the Aquarius haloes,
these massive subhaloes tend to be in the outer parts of their parent halo.  
The subhalo mass fraction above a given
value of ${\rm M_{sub}}/M_{200}$ depends on the mass of the parent
halo. Typically, cluster haloes contain about 25\% more subhalo mass than
galaxy haloes. The Aquarius galactic haloes resolve substructure with
relative mass as small as $10^{-6}$. Subhaloes above this mass contain about
7\% of the total halo mass within $R_{200}$. For clusters, the {\small
  MS-II} resolves subhaloes with relative mass just below
$10^{-5}$. Subhaloes more massive than this already amount to about
7\% of the mass within $R_{200}$. Note, however, that there is a rather large
variance from halo to halo in the subhalo mass fraction above a given
value of ${\rm M_{sub}}/M_{200}$, particularly at the high mass end,
where much of the subhalo mass is typically contributed by one or two
objects. 

\subsection{Redshift dependence}
\begin{figure} 
\resizebox{9cm}{!}{\includegraphics{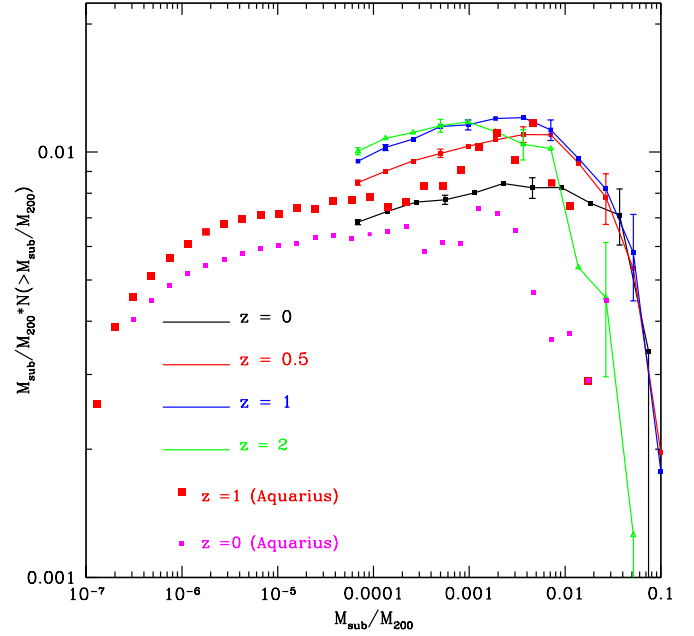}}
\caption{The dependence of the subhalo mass function on redshift. 
The lines show the averaged cumulative subhalo mass functions
for {\small MS-II} haloes in the mass range $[1-3] \times
10^{13}$\solarmass\ at redshift $z=0$ (black), $z=0.5$ (red), $z=1$
(blue) and $z=2$ (green). The filled squares show results for the
Aquarius haloes at $z=1$ (large red) and $z=0$ (magenta). The error
bars on selected points show the error on the mean.}
\label{fig:redshift}
\end{figure}

We expect haloes of a {\em given mass} to contain more subhaloes at
earlier times because the earlier counterparts are both less
concentrated and dynamically younger. This trend was seen in the
simulations of \cite{g04a}. To investigate the redshift dependence of
the subhalo mass function in our simulations, we use a sample of group
haloes of mass $[1, 3] \times 10^{13}$\solarmass\ in the {\small
MS-II}. We also use the Aquarius galaxy haloes at $z=0$ and
their main progenitors at $z=1$,  the mean mass of these halo is
$M_{200} = 1.19 \times 10^{12}$\solarmass\ at $z=0$ and $M_{200} =
6.73 \times 10^{11}$\solarmass at $z=1$.

\begin{figure*} 
\resizebox{8cm}{!}{\includegraphics{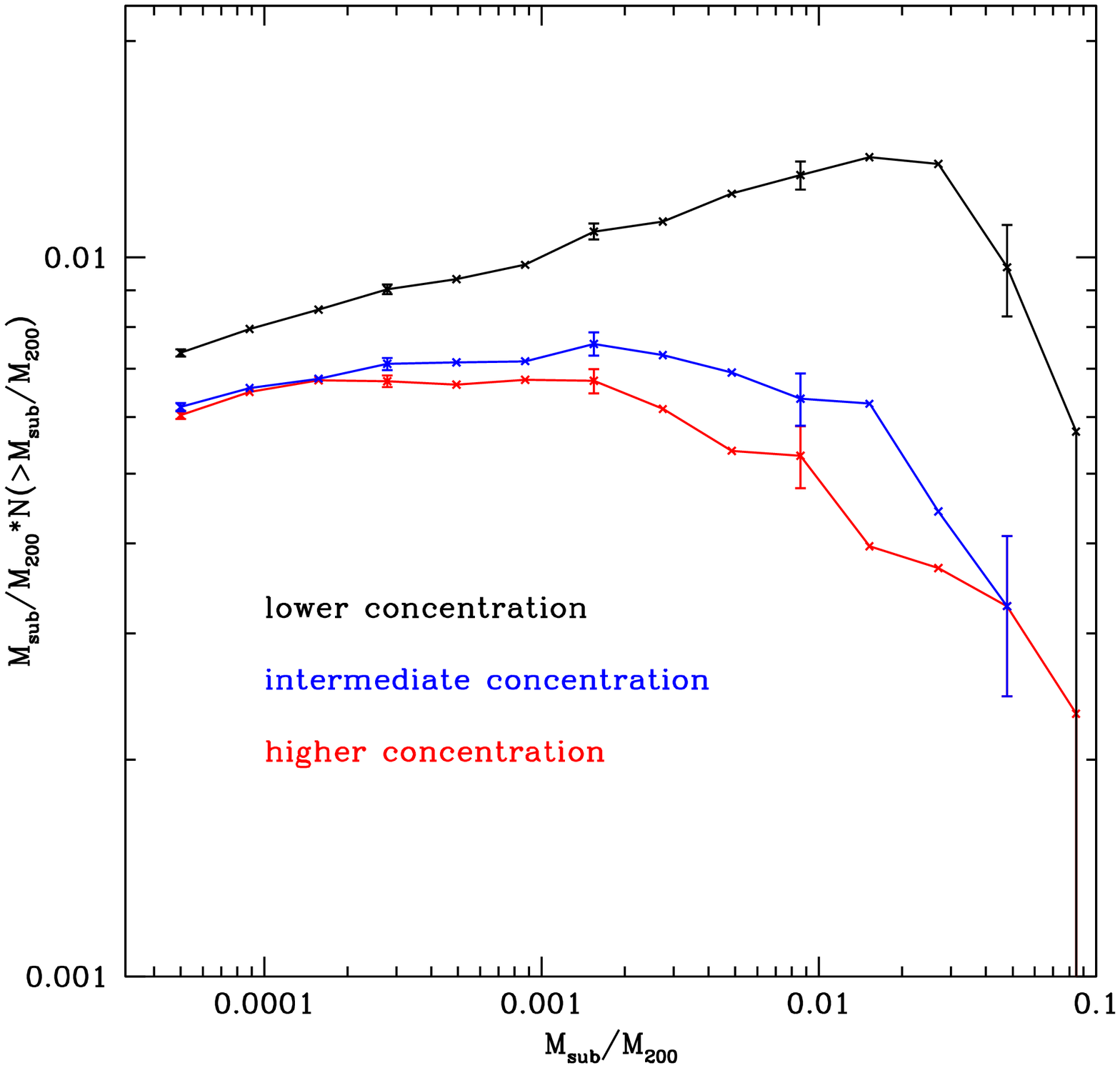}}
\resizebox{8cm}{!}{\includegraphics{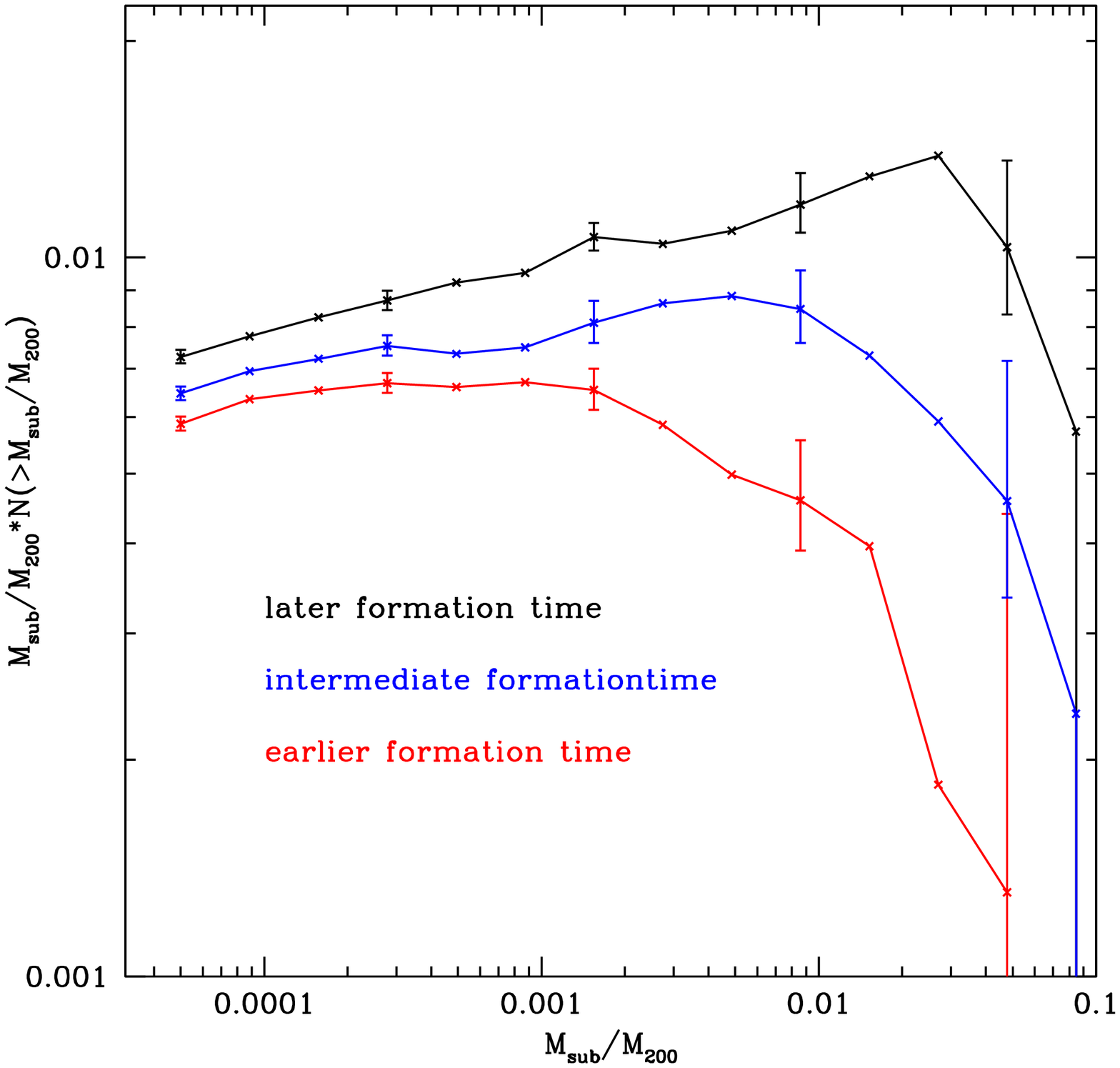}}
\caption{The dependence of the subhalo mass function on the properties
of the parent halo. The left panel shows the dependence on the
concentration parameter of the halo and the right panel on the
formation redshift. The lines show the mean of the cumulative subhalo
mass functions for 219 {\small MS-II} haloes in the mass range $[1-3]
\times 10^{13}h^{-1}M_\odot$. In the two panels, the red solid lines
show the averaged subhalo abundance function of the top third of the
sample, that with the highest concentration parameter (left), and
earliest formation redshift (right). The blue lines correspond to the
intermediate third of the sample and the black lines to the bottom
third, with the lowest concentration parameter and the latest
formation redshift. The error bars on the selected points show
the error on the mean. }
\label{fig:concentration}
\end{figure*}

The results are displayed in Fig~\ref{fig:redshift}. In the {\small MS-II}
sample, there are a total of $219$, $204$, $169$ and $41$ haloes at the four
redshifts shown, $z=0,~0.5,~1,~2$. Clearly, the subhalo mass function evolves
with redshift weakly, but systematically, in the expected sense: at high
redshift haloes of a given mass contain more subhaloes. For example, the
subhalo abundance in groups at $z=0$ is typically $18\%$ lower than at $z=0.5$,
$25\%$ lower than at $z=1$, and $30\%$ lower than at $z=2$. The subhalo mass
function of galaxy haloes evolves slightly more slowly than this, reflecting
the earlier formation epoch of galaxy haloes. In this case the abundance is
only 15\% higher at $z=1$ than at $z=0$.  We note that using
eqn.~(\ref{eq:fits}) with $a$ and $b$ fixed to $-0.94$ and 1.2, respectively,
fits the data shown at each redshift very well. The corresponding parameters are:
\begin{eqnarray}
  \label{eq:fitzz}
  z=0.0\!: \muone=0.0092, \, \mu_{\rm cut}=0.07  \nonumber\\
  z=0.5\!: \muone=0.0118, \, \mu_{\rm cut}=0.06  \nonumber\\
  z=1.0\!: \muone=0.0130, \, \mu_{\rm cut}=0.05   \nonumber\\
  z=2.0\!: \muone=0.0140, \, \mu_{\rm cut}= 0.02   
\end{eqnarray}
For the galaxy haloes the trend persists down
to the smallest subhalo masses resolved in the set of six Aquarius
simulations, ${\rm M_{sub}/M_{200}}=10^{-6}$.

\begin{table*}
\caption{The standard deviation from the mean of the subhalo mass function
at a given fractional mass, $\mmsub=0.01, 0.001, 0.0001$, in units of
the mean.}
\centering
\begin{tabular}{lccccccc}
\hline\hline
$\mu$ &high c & intermediate c & low c & high $z$ & intermediate $z$
  & low $z$ &whole sample\\
\hline
0.01   & 1.52  & 1.23 & 0.67 & 1.5  & 1.01   &0.81 &1.08\\
0.001  & 0.47 & 0.44 & 0.32 & 0.48 & 0.43  &0.33 &0.48 \\
0.0001 & 0.21 & 0.21 & 0.18 & 0.20 & 0.20  &0.20 &0.22 \\
\hline
\end{tabular}
\label{tab:prop}
\end{table*}

\subsection{Dependence on host halo properties}

We now examine how the subhalo mass function depends on two basic
properties of the parent halo: the concentration parameter and
the formation redshift. We also consider whether any such
dependence contributes to the scatter in the subhalo mass function
seen in previous figures, as suggested by~\cite{zentner05}. Earlier
studies have shown that, at least for relatively massive haloes, the
subhalo abundance does depend on the concentration and formation
redshift of the parent halo \citep{g04a,zentner05,shaw06}. With the
{\small MS-II}, we can re-examine the dependence of
subhalo abundance on the properties of the parent halo with much
better statistics than was possible before and also explore the low
mass end of the subhalo mass distribution.

We select 219 haloes in the {\small MS-II} in the mass range $[1,3]
\times 10^{13}$\solarmass, and subdivide this sample into three equal
size subsamples ranked according to concentration parameter or
formation redshift. We evaluate the concentration parameter of a halo
as $V_{\rm max}/V_{200}$ (as was done in~\citealt{g04a}) and we define
its formation redshift as the time when half the mass was
assembled. The results shown in Fig.~\ref{fig:concentration} confirm
the conclusion of~\cite{g04a} that the subhalo abundance decreases
with increasing parent halo concentration and formation redshift. At
the low subhalo mass end, the third least concentrated and most
recently formed haloes have about 25\% more substructures than the
most concentrated and earliest forming third. At the high subhalo mass
end, $M_{\rm sub}/M_{\rm 200} >0.01$, the difference between the two
extreme thirds is about a factor of $2$, significantly larger than at
the lower subhalo mass end.

The variance around the mean of the subhalo mass function for each
subsample and for the sample as a whole are listed in
Table~\ref{tab:prop}, in units of the mean value. The size of the
trends seen in Fig.~\ref{fig:concentration} is comparable to the
scatter in the relations. Nevertheless, the dependence of subhalo
abundance on the concentration and formation time of the parent halo,
at a fixed host halo mass, is much stronger than the dependence on
host halo mass seen in Fig.~\ref{fig:massdep}.

The data presented in Table~\ref{tab:prop} show that the scatter in subhalo
abundance among haloes chosen to have a narrow range of concentration parameters
or formation times is not significantly smaller than the scatter for the halo
population as a whole. Thus, neither concentration nor formation time appear to
be the primary halo property responsible for the observed scatter in the
subhalo mass function.

\section{Conclusion}
We have used a new set of very large cosmological $N$-body
simulations to investigate the statistics of subhalo abundance in
$\Lambda$CDM dark matter haloes. Our results may be summarized as
follows:

\begin{enumerate}

\item The subhalo abundance function of dark matter haloes can be
  well fitted with the functional form proposed by \cite{mike}, which
  is a power-law at the low subhalo mass end and has an exponential
  cut-off at the high mass end, independently of host halo mass and
  redshift.
 
\item The subhalo abundance function depends weakly on host halo
  mass. The difference between typical cluster and galaxy size haloes 
  is about 25 percent at the lower subhalo mass end, substantially weaker
  than expected from previous studies. The scatter in the subhalo
  abundance function of different haloes is larger than the
  systematic trend. 

\item For a given mass halo, the subhalo abundance evolves
  systematically with redshift. The evolution is largest at lower
  redshift ($\sim 18$ percent between $z=0$ and $z=0.5$) and becomes
  very small at high redshift (a few percent between $z=1$ and
  $z=2$). Over the range $z=0-2$, the substructure abundance increases
  by 30\% for haloes corresponding to bright galaxies and poor galaxy
  groups.

\item At fixed mass, dark matter haloes with higher 
  concentration parameters and earlier formation redshifts contain
  fewer subhaloes. The difference is a factor of $25$ percent between
  the top and bottom thirds of the population ranked by concentration
  parameter or formation redshift. However, this dependence on
  concentration and formation redshift is not enough to explain the
  scatter in the subhalo mass function, suggesting that these are not
  the dominant parameters determining the subhalo mass fraction.
\end{enumerate}

\section*{acknowledgements}
The Millennium Simulation and the Millennium-II Simulation were
carried out as part of the programme of the Virgo Consortium on the
Regatta and VIP supercomputers at the Computing Centre of the
Max-Planck Society in Garching. The h{\tt MS} simulation was carried
out using the Cosmology Machine at Durham. LG acknowledges support
from the one-hundred-talents program of the Chinese academy of science
(CAS), the National basic research program of China (program 973 under
grant No. 2009CB24901), {\small NSFC} grants (Nos. 10973018) and an
STFC Advanced Fellowship, as well as the hospitality of the Institute
for Computational Cosmology in Durham, UK. CSF acknowledges a Royal
Society Wolfson Research Merit Award. This work was supported in part
by an STFC rolling grant to the ICC.

\end{document}